\documentclass[final,english]{bullsrsl}[2022/06/15]

\usepackage[latin1]{inputenc}
\usepackage[T1]{fontenc}

\usepackage{natbib} 
\usepackage{graphicx}

\begin{document}
\title{Assembly and testing of Ground Layer Adaptive Optics (GLAO) for ARIES Telescopes}

\author[affil={1}, corresponding]{Purvi}{Udhwani}
\author[affil={1,2}]{Amitesh}{Omar}
\author[affil={1}]{Krishna}{Reddy}

\affiliation[1]{Aryabhatta Research Institute of Observational Sciences, Nainital, UK 263002, India}
\affiliation[2]{Indian Institute of Technology, Kanpur, 208016, India}
\correspondance{udhwanip1@gmail.com}
\date{14th May 2023}
\maketitle


%

\begin{abstract}
This project is focused on evaluating the slowly-varying ground layer seeing component at the optical telescopes of ARIES. To achieve this, we assembled the instrument, consisting of a filter wheel, a CCD camera, and a tip-tilt enabled transparent glass plate integrated within an off-the-shelf unit termed as the AO (Adaptive Optics) unit. The instrument developed by us was deployed on the 1.04-m f/13 Sampurnanand telescope at Manora Peak and the 1.3-m f/4 telescope at Devasthal. This instrument measures the average instantaneous slope (tip/tilt) of the incoming wavefront over the telescope aperture via a fast (within the atmospheric coherence time) sampled image and corrects it via a software-controlled oscillating (tipping/tilting) single thin glass plate. The night observations revealed that the slowly-varying seeing component is significant at both observatories and can be effectively controlled to enhance the sharpness of the celestial images at the two sites. The most significant improvement was measured from 5 arcsec of uncorrected FWHM of a star to 3.4 arcsec of corrected FWHM in the 1.04-m telescope in the evening hours. 
\end{abstract}


\section{Introduction} \label{s:intro}
\begin{figure}[t]
\centering
\includegraphics[width=\textwidth]{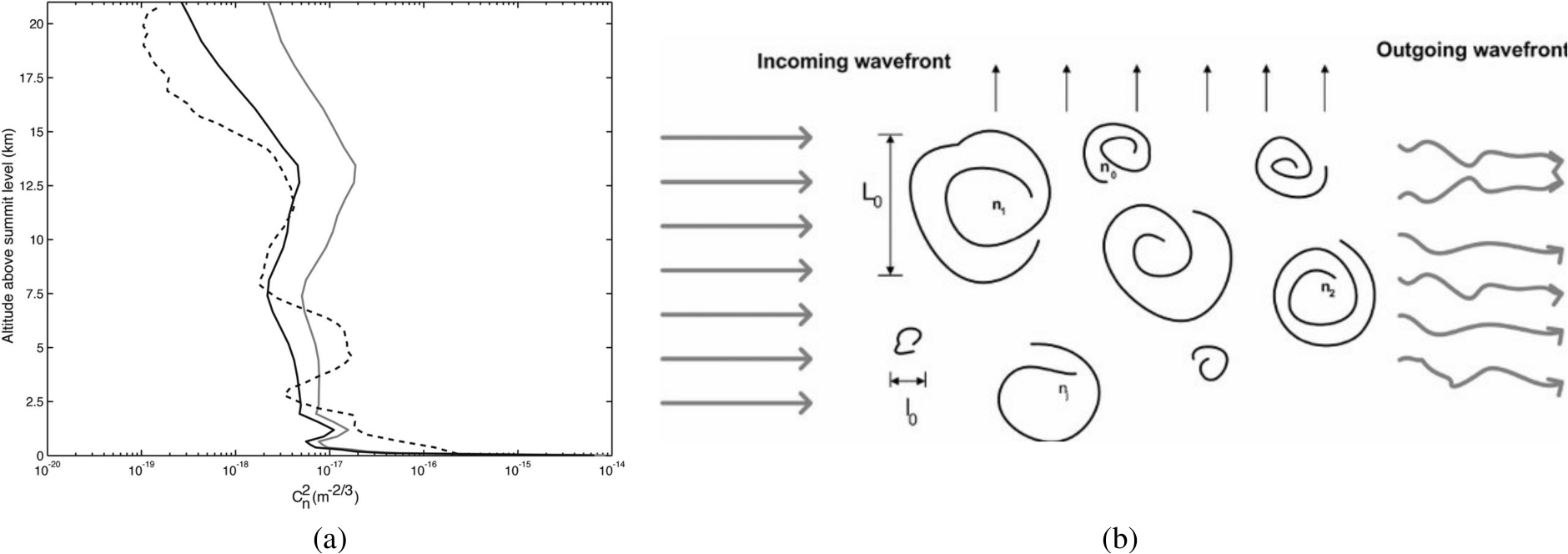}
\smallskip

\begin{minipage}{12cm}
\caption{(a) Turbulence profile over different altitudes depicting how the refractive index structure function varies within different layers of the atmosphere \citep{AnotherLookattheRefractiveIndexStructureFunction}.
(b) Changes in the wind due to refractive index variations called eddies of different scales, act like little lenses in a layer for the wavefront (Chapter 11-\citet{inbook}).}
\label{fig:eddies}
\end{minipage}
\end{figure}

The atmospheric turbulence in optical astronomy is a critical limiting factor for the quality and sensitivity of ground-based optical observations and poses significant challenges to astronomers. The primary consequence of turbulence is the degradation of image quality, commonly known as ''seeing".  This 'seeing' caused by the atmosphere is due to random refractive index fluctuations within its turbulent layers, caused by the interaction between the Earth's surface and the overlying atmosphere, surface features such as buildings, trees, and uneven terrain disrupt the smooth flow of air \citep{10.1117/12.566235}, also called Kolmogorov turbulence depicted by a 'two-thirds power law' equation.
\begin{center}
\begin{equation}
    D_n (\Delta r) = C_{N} ^2 (z) \Delta r ^{2/3}   \hspace{10mm} l_0 < \Delta r < L_0
    \label{eq:kolmogorov}
\end{equation}
 \end{center} 
The primary function of an optical telescope is to effectively collect light photons from distant sources and provide adequate angular resolution to spatially resolve sources in the sky. The larger the diameter of the primary mirror, the better its light-gathering capability and diffraction limit (angular resolution). The signal-to-noise ratio in the deep long-integration images is directly proportional to the effective collecting area of the telescope and inversely proportional to the solid angle formed by the seeing disk or diffraction limit of the telescope whichever is higher.
This turbulence impacts the wavefront in various ways like high spatial-frequency beam spreading and low spatial-frequency beam-wander along with some intensity variations \citep{doi:10.1080/00107514.2017.1344320}.
A measure of the strength of this turbulence is the refractive index structure function or $C_{N} ^2$ as mentioned in \ref{eq:kolmogorov} which is specific to locations and at each location varies differently with altitude. (Fig.\,\ref{fig:eddies}(a))

\begin{figure}[t]
\centering
\includegraphics[width=\textwidth]{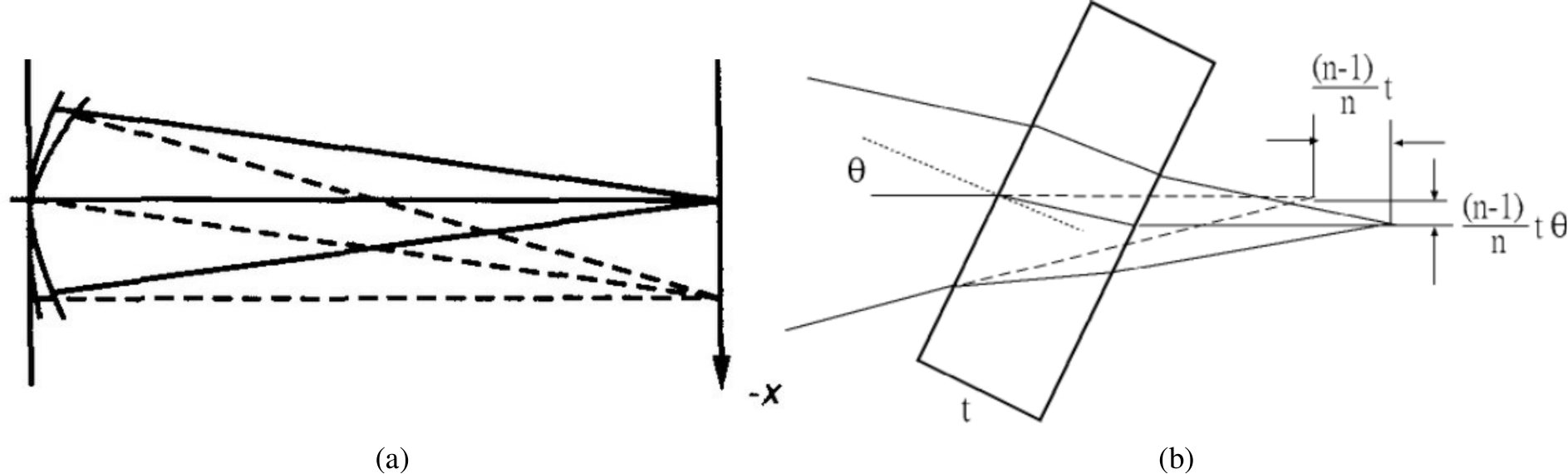}
\smallskip

\begin{minipage}{12cm}
\caption{(a) The tilting of the wavefront induces a centroid shift in the image. The control algorithm of the tip-tilt system utilizes this image shift to accurately calculate the glass plate's precise movement \citep{1992aooe...11....2W}.
(b) The compensation for wavefront tilt involves the movement of a plane parallel plate, which is determined by the plate's thickness (t), the angle ($\theta$) between the optical axis and the plate's normal, and the refractive index (n) of the material.}
\label{fig:centroid}
\end{minipage}
\end{figure}

Due to atmospheric turbulence, starlight passing through the Earth's atmosphere gets distorted, causing the stars to twinkle and reducing the clarity of the images. This distortion causes the incoming wavefront to be tilted and hence a shift in the centroid of the image. By using a tip-tilt system, these disturbances can be compensated for in real-time, resulting in sharper images. This technique is particularly effective for reducing low-frequency vibrations and atmospheric blurring.
In the simplest form, an average instantaneous slope of the incoming wavefront's tip/tilt can be determined from the centroid of a fast (within the coherence time) sampled image. This is equivalent to a single-element wavefront sensor while a multi-element wavefront sensor (e.g., Shack-Hartmann) determines wavefront slopes at some spatial resolution over the full aperture of the telescope. This single-element or average instantaneous slope of the incoming wavefront's tip/tilt can be compensated using a thin glass plate in transmission optical systems such as simple CCD imagers on the telescope. The schematics are shown in Fig.\,\ref{fig:centroid}. The measurement is done using the following set of formulas:
\begin{equation}\label{eq:plate}
    \Delta y = \frac{t \theta (n-1)}{n}
\end{equation}
\begin{equation}\label{eq:converge}
    \Delta x = \frac{(n-1)}n{t}
\end{equation}
Where $t$ is the thickness of the plane parallel plate, $n$ is the refractive index of the material, $\theta$ is the angle made by the normal of the plate with the optical axis. Equation \ref{eq:plate} depicts the linear deviation in the y direction caused by the tilting of the plane-parallel plate and Equation \ref{eq:converge} is the linear deviation in the x direction caused by converging of the beam. As shown in Fig.\,\ref{fig:centroid}(b), in a real case scenario, both effects combine to cause linear deviation in the x and y directions

In this project, we aim to correct the low-order beam wander caused by the dynamic tilt of the wavefront or "tip-tilt" correction. Section\,\ref{s:algo} describes the setup of the instrument used to do so and the methodology. The results are described in Section\,\ref{s:results}.

\section{Instrumental setup and Methodology}\label{s:algo}
This AO unit used in this work was constructed using three main components from SBIG or Santa Barbara Instrument Group, a company that specializes in producing astronomical instruments, particularly CCD (Charge-Coupled Device) cameras and related equipment. The components used in this setup and their length/distance from the top to the last optical surface along with other features are listed in Table\,\ref{tab:setup}.

\begin{itemize}
    \item  Filter Wheel: The Filter Wheel is equipped with L, R, B, and G filters- occupying 4 out of the 8 filter slots, a pick-off mirror, and a self-guiding CCD. A pick-off mirror placed in front of the filters directs the light from the guide star onto the guiding CCD to obtain correction parameters.
    \item  ST-8300 CCD Camera: The ST-8300 CCD camera is primarily used for imaging purposes to capture the source light. It also has an optional remote guide head that houses a small guiding CCD similar to the filter wheel CCD. However, this remote guide head is not utilized as the light from the guide star would be attenuated through the filters.
    \item  AO-8 System: The AO-8 system features a tip-tilt mirror with two geared stepper motors for movement along the X and Y directions. These motors enable the tipping and tilting of a 6mm thick plane parallel plate, which transmits the incoming source light to the imaging CCD. 
\end{itemize} 

\begin{table}
\centering
\begin{minipage}{120mm}
\caption{Components utilized in the system and their corresponding optical length. This parameter was employed to fine-tune the focus relative to the back focal length of the telescope up to the image plane on the CCD camera. The difference between the total optical length of the instrument and the back focal length of the telescope determines the necessary size of the supporting studs.}
\label{tab:setup}
\end{minipage}
\bigskip

\begin{tabular}{ll}
\hline
\textbf{Component} & \textbf{Length (optical) and features} \\
\hline
Filter Wheel & b=2.128 inch, pixel size=7.4 $\mu$m \\
ST-8300 CCD  & b=0.69 inch, pixel size= 5.4 $\mu$m \\ 
AO-8T        & b=2.02 inch, tip rate=18.74 $\deg s^{-1}$, tilt$_{xy}$=$\pm 9.6 \deg$\\
\hline
\end{tabular}
\end{table}

\begin{figure}[p]
\centering    
\includegraphics[width=0.75\linewidth]{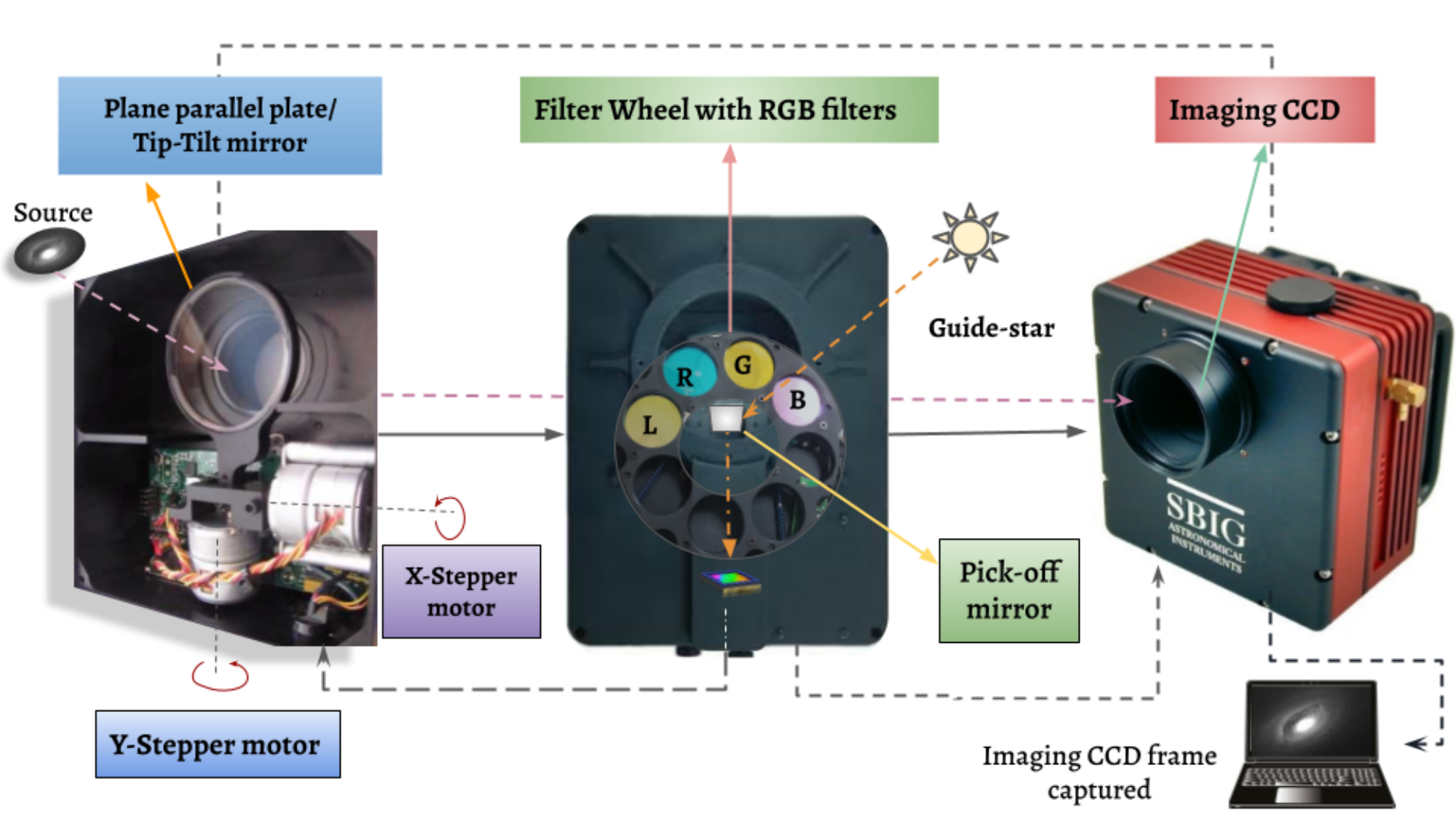}
\smallskip

\begin{minipage}{12cm}
\caption{The complete AO setup - The light path of the source and guide star is represented by colored dashed arrows (pink and yellow). The sub-components are labeled within the colored boxes, while the electrical connections are depicted as grey dashed lines. The black dashed line indicates the USB connection linking the imaging CCD to the computer, responsible for converting visual information into digital data. }
\label{fig:setup}
\end{minipage}
\end{figure}

After assembling these components (as depicted in Fig.\,\ref{fig:setup}) in the laboratory, to assess the performance of the instrument setup, a laboratory-based test was conducted using low-power lasers. The purpose of this experiment was twofold - first, to investigate the effects of different filters on the source light, and second, to determine the optimal focus parameters for the small CCD.
Furthermore, the instrument setup was tested first on small aperture telescopes to assess the PSF at various telescope sites.

\begin{figure}[p]
\centering
\includegraphics[width=\textwidth]{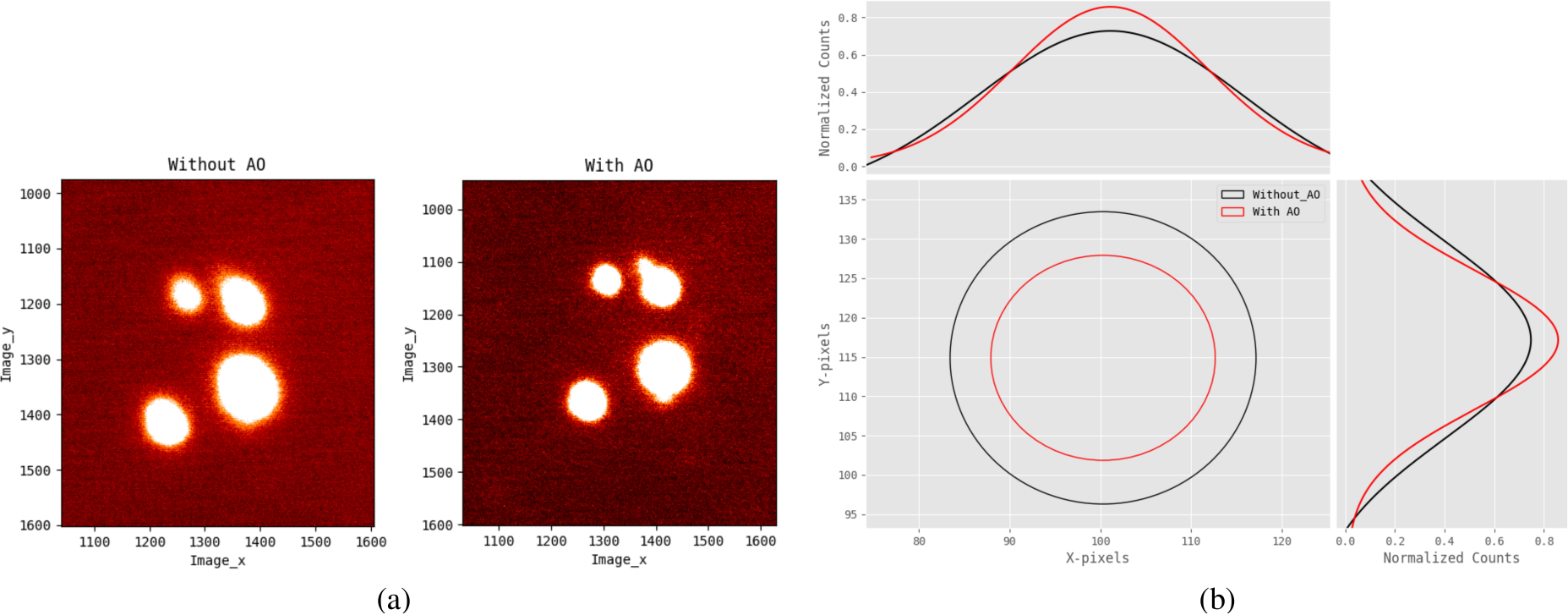}
\smallskip

\begin{minipage}{12cm}
\caption{(a) An HII region, observed with the 1.04-m telescope on  January 8, 2023, in the late evening hours. The source on the top right seems like a single extended source, while in the AO-corrected image on the right, two partially resolved stars can be seen.
(b) The 2-D Gaussian fit PSF for one of the sources in the cluster and the corresponding 1-D X and Y Gaussian PSF fit for the same.}
\label{fig:results}
\end{minipage}
\end{figure}

\section{Results and Summary}\label{s:results}

After conducting some tests on smaller telescopes at ARIES and acquiring a comprehensive understanding of the assembled AO instrument, the unit was deployed and evaluated at two observing sites: the 1.04-m Sampurnanand telescope (F=13520, f/13, plate scale=0.08"/pixel) located at the Manora peak (alt. = 1950 m), Nainital campus and the 1.3-m telescope(F=5150, f/4, plate scale=0.22"/pixel) situated at the Devasthal campus (alt. = 2450 m). By testing the instrument at these two sites, having different altitudes, and seeing conditions, valuable insights were obtained regarding the performance of the AO instrument assembled by us.
Correction of the tip-tilt alone can substantially enhance the quality of the image (Fig.\,\ref{fig:results}). The system developed and assembled by us is a straightforward, cost-effective, and efficient adaptive optics system designed to elucidate the impact of turbulence on image quality at ARIES's telescope sites. The current results demonstrate a notable improvement in the FWHM of the star images for various sources by using the til-tilt AO unit.
The following results were obtained after the analysis of the project:
\begin{itemize}
    \item Through extensive testing of various filters and sampling times, we found that the red filter provided the most significant improvement in seeing conditions, regardless of whether or not adaptive optics (AO) corrections were applied. 
    \item A comparison between the results obtained from the 1.04-m and 1.3m telescopes revealed marginally better image (seeing) improvements at the 1.04-m telescope, possibly due to the higher amplitude of ground-layer turbulence at Manora Peak compared to that at Devasthal.
    \item The corrections were stable for long-exposure images as well.
\end{itemize}

\begin{acknowledgments}
The authors would like to thank ARIES for providing the observational facilities (Sampurnanand 40-inch and the 1.3m telescope) and computational facilities necessary to obtain the data for testing this AO unit. We would also thank the BINA project for giving us the opportunity and support to present our poster in the consortium and to write this manuscript for the proceedings.
\end{acknowledgments}

\begin{furtherinformation}

\begin{orcids}
\orcid{0009-0007-3947-5838}{Purvi}{Udhwani}
\orcid{0000-0002-7449-6593}{Amitesh}{Omar}
\end{orcids}

\begin{authorcontributions}
Purvi Udhwani (PU) assembled the components in the lab under the supervision of Dr. Amitesh Omar and conducted night observations with the help of Krishna Reddy. Apart from supervision, Dr. Amitesh Omar also conceptualized the project and provided a methodology for this work. Krishna Reddy monitored data curation, validation, and visualization of the method. The writing of this original draft and investigation of the work was done by PU.
\end{authorcontributions}

\begin{conflictsofinterest}
The authors declare no conflict of interest.
\end{conflictsofinterest}

\end{furtherinformation}

\bibliographystyle{bullsrsl-en}

\bibliography{S02-P07_UdhwaniP}

\end{document}